\documentclass[floatfix,twocolumn,showpacs,aps,tightenlines]{revtex4-1}
\usepackage{graphicx}
\usepackage{color}
\usepackage{psfrag}
\usepackage{dcolumn}
\usepackage{bm}

\usepackage{amssymb}
\usepackage{amsmath}
\usepackage{amsfonts}
\usepackage{longtable}
\usepackage{xspace}

\newcommand{\dmt}{\delta{m}}

\newcommand{\beq}{\begin{equation}}
\newcommand{\eeq}{\end{equation}}

\newcommand{\be}{\begin{equation}}
\newcommand{\ee}{\end{equation}}
\newcommand{\bea}{\begin{eqnarray}}
\newcommand{\eea}{\end{eqnarray}}
\newcommand{\bes}{\begin{subequations}}
\newcommand{\ees}{\end{subequations}}

\usepackage{bm}

\begin{document}

\title{The nonspinning binary black hole merger scenario revisited}

\author{James Healy}
\author{Carlos O. Lousto}
\author{Yosef Zlochower}
\affiliation{Center for Computational Relativity and Gravitation,
School of Mathematical Sciences,
Rochester Institute of Technology, 85 Lomb Memorial Drive, Rochester,
New York 14623}

\date{\today}

\begin{abstract}
We present the results of 14 simulations of nonspinning
black hole binaries with mass ratios $q=m_1/m_2$ in the range
$1/100\leq q\leq1$. For each of these simulations we perform three runs
at increasing resolution to assess the finite difference
errors and to extrapolate the results to infinite resolution.
For $q\geq 1/6$, we follow the evolution of the binary typically for the last
ten orbits prior to merger.
By fitting the results of these simulations, we accurately model the peak
luminosity, peak waveform frequency and amplitude, 
and the recoil of the remnant hole for unequal mass 
nonspinning binaries. We verify the accuracy of these new models and
compare them to previously
existing empirical formulas.
These new fits provide a basis for a hierarchical approach to produce
more accurate remnant formulas in the generic precessing case.
They also provide input to gravitational waveform
modeling.
\end{abstract}

 \maketitle

\section{Introduction}\label{sec:Intro}

Recent LIGO observations
\cite{Abbott:2016blz,Abbott:2016nmj, TheLIGOScientific:2016pea}
of gravitational waves agree with the predictions based on
supercomputer simulations \cite{Pretorius:2005gq, Campanelli:2005dd, Baker:2005vv}
of the merger of binary black holes.
Direct comparison of the first observed signal, GW150914, with 
targeted numerical relativity waveforms have been performed in
\cite{Abbott:2016blz,Abbott:2016apu,Lovelace:2016uwp}. This allows the
study of their astrophysical properties, such as masses, spins
and location in the universe~\cite{TheLIGOScientific:2016pea}.

The breakthroughs~\cite{Pretorius:2005gq,Campanelli:2005dd,Baker:2005vv}
in numerical relativity allowed for not only the detailed predictions for the
gravitational waves from the late inspiral, plunge, merger and
ringdown of black hole binary systems (BHB)
\cite{Mroue:2013xna, Husa:2015iqa, Jani:2016wkt, Healy:2017psd}, but also
for determining how the 
individual masses and spins of the orbiting binary relate to the
properties of the final remnant
black hole produced after merger. This relationship~\cite{Healy:2014yta}
can be used as a consistency check for the observations of the inspiral
and, independently, the merger-ringdown signals as tests of general relativity
\cite{Ghosh:2016qgn,TheLIGOScientific:2016src,TheLIGOScientific:2016pea}.

In Ref.~\cite{Healy:2016lce} we revisited the scenario of 
aligned-spin BHB mergers we first studied in~\cite{Healy:2014yta}.
There we added 71 new simulations to our
original 36 to verify and improve the
fitting formulas that related the aligned
spin binaries initial parameters [mass ratio and intrinsic spins along
the orbital angular momentum for black holes 1 and 2
$(q,\alpha_1^L,\alpha_2^L)$] to
the final black hole mass, spin and recoil
$(m_f,\alpha_f,V_f)$. We have also modeled in \cite{Healy:2016lce}
the peak luminosity produced by the binary merger, as this is of
astrophysical and gravitational wave observations 
interest~\cite{Abbott:2016blz, Abbott:2016nmj, TheLIGOScientific:2016pea, Keitel:2016krm}.
In this paper we introduce a model for the gravitational wave
frequency and amplitude at the peak of the strain $(2,2)$ mode.

While the modeling of the final mass and spin by \cite{Healy:2016lce}
has proven to be extremely accurate, with estimated errors of the order
of $0.1\%$ and $0.2\%$ respectively, the recoil velocities and
peak luminosity typical errors are of the order of $5\%$.
This is because we are able to use the 
the final isolated horizon measures for the mass and spin~\cite{Dreyer:2002mx}
in the fittings,
while the recoil (or radiated linear momentum)
and peak luminosity are directly measured from the waveforms.
In typical BHB simulations waveform accuracy is mostly affected by
the finite extraction radius, finite difference of the numerical integration
method and finite number of extracted radiation multipoles
(See appendices of \cite{Healy:2014yta,Healy:2016lce}).
In this paper we will improve on the finite difference errors by 
computing each new simulation with three resolutions, labeled as
N100, N120, N140 (characterizing the increasing
number of gridpoints in the innermost refinement level of the adaptive
mesh refinement grid hierarchy). The three existing simulations were performed
at equivalent resolutions of N100, N144, and N173 for $q=1/10$, 
N144, N173, and N207 for $q=1/15$, and N100, N144, and N207 for 
$q=1/100$.
We also use a proven method to perturbatively extrapolate 
the results from a finite distance observer location to infinity
\cite{Nakano:2015pta}, and include up to 
$\ell=6$ multipoles in the computation of the radiative quantities.

The paper is organized as follows. In Sec. \ref{sec:FN} we
describe the methods and criteria for producing the new
simulations.
We next study in Sec.~\ref{sec:rvs0} the computation and modeling
of the recoil velocities of the remnant of the merger of two nonspinning
black holes. In Sec.~\ref{sec:pls0} we use the simulations and its
extrapolations to model the peak luminosities
and compare them to recent fits.
In Sec~\ref{sec:oms0} we propose expansions and fit 
the waveform frequency and amplitude at the peak of the strain
mode $(2,2)$.
We conclude with a discussion in 
Sec.~\ref{sec:Discussion} of the use and potential extensions 
of this work to spinning and precessing binaries as well as
the gravitational waveform modeling.

\section{Full Numerical Evolutions}\label{sec:FN}

In order to make systematic studies and build a data bank of
full numerical simulations, it is crucial to develop efficient
numerical algorithms, since large computational resources are required.

We evolve the BHB data sets using the {\sc
LazEv}~\cite{Zlochower:2005bj} implementation of the moving puncture
approach~\cite{Campanelli:2005dd,Baker:2005vv} with the conformal
function $W=\sqrt{\chi}=\exp(-2\phi)$ suggested by
Ref.~\cite{Marronetti:2007wz}.  For the 11 new runs presented here,
with $1/6\leq q\leq1$, we use
centered, sixth-order finite differencing in
space~\cite{Lousto:2007rj} and a fourth-order Runge Kutta time
integrator (note that we do not upwind the advection terms)
and a 5th-order Kreiss-Oliger dissipation operator.
%

Our code uses the {\sc EinsteinToolkit}~\cite{Loffler:2011ay,
einsteintoolkit} / {\sc Cactus}~\cite{cactus_web} /
{\sc Carpet}~\cite{Schnetter-etal-03b}
infrastructure.  The {\sc
Carpet} mesh refinement driver provides a
``moving boxes'' style of mesh refinement. In this approach, refined
grids of fixed size are arranged about the coordinate centers of both
holes.  The {\sc Carpet} code then moves these fine grids about the
computational domain by following the trajectories of the two BHs.

To compute the initial low eccentricity orbital parameters
we use the post-Newtonian techniques described in~\cite{Healy:2017zqj}.
To compute the numerical initial data, we use the puncture
approach~\cite{Brandt97b} along with the
{\sc TwoPunctures}~\cite{Ansorg:2004ds} code implementation. 

We use {\sc AHFinderDirect}~\cite{Thornburg2003:AH-finding} to locate
apparent horizons.  We measure the magnitude of the horizon spin using
the {\it isolated horizon} (IH) algorithm detailed in
Ref.~\cite{Dreyer02a} and as implemented in Ref.~\cite{Campanelli:2006fy}.
Note that once we have the horizon spin, we can calculate the horizon
mass via the Christodoulou formula 
${m_H} = \sqrt{m_{\rm irr}^2 + S_H^2/(4 m_{\rm irr}^2)}\,,$
where $m_{\rm irr} = \sqrt{A/(16 \pi)}$, $A$ is the surface area of
the horizon, and $S_H$ is the spin angular momentum of the BH (in
units of $M^2$). 
We measure radiated energy,
linear momentum, and angular momentum, in terms of the radiative Weyl
Scalar $\psi_4$, using the formulas provided in
Refs.~\cite{Campanelli:1998jv,Lousto:2007mh},
Eqs. (22)-(24) and (27) respectively. However, rather than
using the full $\psi_4$, we decompose it into $\ell$ and $m$ modes and
solve for the radiated linear momentum, dropping terms with $\ell >
6$.  The formulas in Refs.~\cite{Campanelli:1998jv,Lousto:2007mh} are
valid at $r=\infty$.  We extract the radiated energy-momentum at
finite radius and extrapolate to $r=\infty$. We find that the new
perturbative extrapolation described in Ref.~\cite{Nakano:2015pta} provides the
most accurate waveforms. 

\section{Results}\label{sec:Results}


We perform a set of 11 new runs for nonspinning binaries in the mass
ratio range $1/6\leq q\leq1$ as described in Table \ref{tab:ID0}
and include the $q=1/10$ case reported in \cite{Hinder:2013oqa}
and the $q=1/15$ and $1/100$ cases reported in \cite{Lousto:2010ut,Lousto:2010qx,Nakano:2011pb,Lousto:2010tb}

\begin{table*}
\caption{Initial data parameters for the quasi-circular
configurations with a smaller mass black hole (labeled 1),
and a larger mass black hole (labeled 2). The punctures are located
at $\vec r_1 = (x_1,0,0)$ and $\vec r_2 = (x_2,0,0)$, with momenta
$P=\pm (P_r, P_t,0)$, spins $\vec S_i = (0, 0, 0)$, mass parameters
$m^p/m$, horizon (Christodoulou) masses $m^H/m$, total ADM mass
$M_{\rm ADM}$, and dimensionless spins $a/m_H = S/m_H^2$.
}\label{tab:ID0}
\begin{ruledtabular}
\begin{tabular}{lccccccccccccc}
$q=m_1^H/m_2^H$   & $x_1/m$ & $x_2/m$  & $P_r/m$    & $P_t/m$ & $m^p_1/m$ & $m^p_2/m$ & $S_1/m^2$ & $S_2/m^2$ & $m^H_1/m$ & $m^H_2/m$ & $M_{\rm ADM}/m$ & $a_1/m_1^H$ & $a_2/m_2^H$\\
\hline
0.0100 & -4.95 & 0.05 & -1.03e-5 & 0.00672 & 0.0087 & 0.9896 & 0 & 0 & 0.0099 & 0.9907 & 1.0000 & 0 & 0 \\
0.0667 & -6.86 & 0.44 & -1.60e-4 & 0.02907 & 0.0576 & 0.9362 & 0 & 0 & 0.0625 & 0.9404 & 1.0000 & 0 & 0 \\
0.1000 & -7.63 & 0.75 & -1.69e-4 & 0.03670 & 0.0852 & 0.9074 & 0 & 0 & 0.0913 & 0.9126 & 1.0000 & 0 & 0 \\
0.1667 & -9.00 & 1.50 & -2.19e-4 & 0.04590 & 0.1358 & 0.8511 & 0 & 0 & 0.1429 & 0.8571 & 0.9952 & 0 & 0 \\
0.2000 & -8.96 & 1.79 & -2.55e-4 & 0.05116 & 0.1589 & 0.8266 & 0 & 0 & 0.1667 & 0.8333 & 0.9947 & 0 & 0 \\
0.2500 & -8.80 & 2.20 & -3.08e-4 & 0.05794 & 0.1913 & 0.7923 & 0 & 0 & 0.2000 & 0.8000 & 0.9940 & 0 & 0 \\
0.3333 & -8.44 & 2.81 & -3.83e-4 & 0.06677 & 0.2401 & 0.7411 & 0 & 0 & 0.2500 & 0.7500 & 0.9930 & 0 & 0 \\
0.4000 & -8.04 & 3.21 & -4.50e-4 & 0.07262 & 0.2751 & 0.7045 & 0 & 0 & 0.2857 & 0.7143 & 0.9924 & 0 & 0 \\
0.5000 & -7.33 & 3.67 & -5.72e-4 & 0.08020 & 0.3216 & 0.6557 & 0 & 0 & 0.3333 & 0.6667 & 0.9916 & 0 & 0 \\
0.6000 & -7.19 & 4.31 & -5.46e-4 & 0.08206 & 0.3632 & 0.6138 & 0 & 0 & 0.3750 & 0.6250 & 0.9914 & 0 & 0 \\
0.6667 & -7.05 & 4.70 & -5.29e-4 & 0.08281 & 0.3883 & 0.5887 & 0 & 0 & 0.4000 & 0.6000 & 0.9913 & 0 & 0 \\
0.7500 & -6.29 & 4.71 & -6.86e-4 & 0.08828 & 0.4159 & 0.5591 & 0 & 0 & 0.4286 & 0.5714 & 0.9907 & 0 & 0 \\
0.8500 & -6.49 & 5.51 & -5.29e-4 & 0.08448 & 0.4477 & 0.5290 & 0 & 0 & 0.4595 & 0.5405 & 0.9912 & 0 & 0 \\
1.0000 & -10.00& 10.00& -1.04e-4 & 0.06175 & 0.4930 & 0.4930 & 0 & 0 & 0.5000 & 0.5000 & 0.9943 & 0 & 0 \\
\end{tabular}
\end{ruledtabular}
\end{table*}

The evolution of these 14 nonspinning binaries leads to
recoil velocities, peak luminosities, peak frequency and
peak amplitude 
as shown in Tables \ref{tab:0rem}, \ref{tab:0omvals}, 
and \ref{tab:0ampvals}.  In Tables ~\ref{tab:0omvals} and 
\ref{tab:0ampvals}, we also include the peak frequency
and peak amplitude values calculated from the $(2,2)$ mode of
$\Psi_4$ and the first time derivative of the strain, $N$.

For the recoil velocity and peak luminosity, the error reported 
in Table \ref{tab:0rem} is calculated from the finite resolution
and finite observer location errors.  To estimate the finite resolution
error we determine compare the results of the highest resolution with
those obtained by a Richardson extrapolation of all resolutions.
 To estimate the
finite observer location error, we use the perturbative extrapolation
technique in Ref~\cite{Nakano:2015pta} at all observer locations and
take the difference between the largest and smallest radii.  
Calculating the error in this way overestimates the error,
since as $r_{obs} \rightarrow \infty$ the difference between the values at
successive observers decreases.  Even with this conservative
calculation of the observer location error, the finite
resolution error is typically the dominant error source, but we include
both in the total error estimate by adding both sources in quadrature.

In addition to finite resolution and observer location error, the peak
frequency has another source
of error.  To estimate the peak frequency, we need to interpolate the 
time-series data to find the peak, and since in the region of the peak
amplitude, $d\omega/dt$ is large, this introduces an uncertainty.
To estimate this, we use the value of the frequency at the interpolated
peak, and then the difference between the two nearest time points are
used as the error.  This error is on the order of $0.5-1.0\%$ 
and decreases with increasing resolution.  This third error is added
to the finite observer and resolution error in quadrature and is
quoted as the errors in Table \ref{tab:0omvals}.  For the peak amplitude, 
this type of error is negligible since in the region of the peak,
$dA/dt = 0$, and there are enough data points in the area to model the
peak accurately without interpolation.  Nonetheless, we can calculate the
error from interpolation in the amplitude by taking the difference of the
interpolated value with the nearest data point.  

\begin{table}
\caption{Recoil velocity and peak luminosity
for nonspinning binaries.  Values are extrapolated to infinite resolution 
and infinite observer location and the error reflects the error in both operations added in quadrature.
}
\label{tab:0rem}
\begin{ruledtabular}
\begin{tabular}{crr}
$q$    & $V_{rem}$         & $L_{peak}$ \\
\hline
0.0100 & $   0.87\pm0.04 $ & $ 1.214\times 10^{-6}\pm 5.641\times10^{-9} $ \\
0.0667 & $  33.56\pm0.50 $ & $ 4.417\times 10^{-5}\pm 4.655\times10^{-7} $ \\
0.1000 & $  62.51\pm0.56 $ & $ 9.009\times 10^{-5}\pm 9.736\times10^{-7} $ \\
0.1667 & $ 118.32\pm2.85 $ & $ 2.185\times 10^{-4}\pm 8.209\times10^{-6} $ \\
0.2000 & $ 141.17\pm3.90 $ & $ 2.729\times 10^{-4}\pm 3.193\times10^{-6} $ \\
0.2500 & $ 160.89\pm3.82 $ & $ 3.718\times 10^{-4}\pm 4.820\times10^{-6} $ \\
0.3333 & $ 177.89\pm3.91 $ & $ 5.298\times 10^{-4}\pm 5.389\times10^{-6} $ \\
0.4000 & $ 173.55\pm3.52 $ & $ 6.358\times 10^{-4}\pm 5.939\times10^{-6} $ \\
0.5000 & $ 154.82\pm2.94 $ & $ 7.775\times 10^{-4}\pm 6.944\times10^{-6} $ \\
0.6000 & $ 126.04\pm2.28 $ & $ 8.809\times 10^{-4}\pm 8.674\times10^{-6} $ \\
0.6668 & $ 102.29\pm1.55 $ & $ 9.296\times 10^{-4}\pm 9.733\times10^{-6} $ \\
0.7500 & $  76.15\pm1.56 $ & $ 9.749\times 10^{-4}\pm 8.184\times10^{-6} $ \\
0.8500 & $  43.23\pm0.59 $ & $ 1.010\times 10^{-3}\pm 1.074\times10^{-5} $ \\
1.0000 & $   0.00\pm0.00 $ & $ 1.038\times 10^{-3}\pm 3.739\times10^{-5} $ \\
\end{tabular}
\end{ruledtabular}
\end{table}

\begin{table*}
\caption{ Peak frequency of the 22 mode measured from the strain, news, and
$\Psi_4$.  Values are extrapolated to infinite observer and resolution, and
error values take into account both operations, plus the additional error 
introduced by finding the peak of the waveform, all added in quadrature.
}
\label{tab:0omvals}
\begin{ruledtabular}
\begin{tabular}{cccc}
$q$    & $m\omega_{22}^{H\mathrm{peak}}$ & $m\omega_{22}^{N\mathrm{peak}}$ & $m\omega_{22}^{\Psi_4\mathrm{peak}}$ \\
\hline
0.0100 & $ 0.2825 \pm 0.0007 $ & $ 0.3303 \pm 0.0025 $ & $ 0.3407 \pm 0.0152 $ \\
0.0667 & $ 0.2904 \pm 0.0008 $ & $ 0.3468 \pm 0.0015 $ & $ 0.3785 \pm 0.0041 $ \\
0.1000 & $ 0.2947 \pm 0.0034 $ & $ 0.3586 \pm 0.0011 $ & $ 0.3955 \pm 0.0020 $ \\
0.1667 & $ 0.3097 \pm 0.0028 $ & $ 0.3912 \pm 0.0138 $ & $ 0.4061 \pm 0.0096 $ \\
0.2000 & $ 0.3153 \pm 0.0021 $ & $ 0.3757 \pm 0.0060 $ & $ 0.4203 \pm 0.0045 $ \\
0.2500 & $ 0.3208 \pm 0.0022 $ & $ 0.3920 \pm 0.0024 $ & $ 0.4307 \pm 0.0075 $ \\
0.3333 & $ 0.3323 \pm 0.0024 $ & $ 0.4097 \pm 0.0027 $ & $ 0.4467 \pm 0.0018 $ \\
0.4000 & $ 0.3384 \pm 0.0024 $ & $ 0.4125 \pm 0.0034 $ & $ 0.4693 \pm 0.0111 $ \\
0.5000 & $ 0.3463 \pm 0.0026 $ & $ 0.4285 \pm 0.0027 $ & $ 0.4675 \pm 0.0015 $ \\
0.6000 & $ 0.3517 \pm 0.0027 $ & $ 0.4364 \pm 0.0027 $ & $ 0.4786 \pm 0.0018 $ \\
0.6667 & $ 0.3512 \pm 0.0030 $ & $ 0.4401 \pm 0.0028 $ & $ 0.4959 \pm 0.0113 $ \\
0.7500 & $ 0.3566 \pm 0.0028 $ & $ 0.4430 \pm 0.0026 $ & $ 0.4924 \pm 0.0050 $ \\
0.8500 & $ 0.3565 \pm 0.0029 $ & $ 0.4427 \pm 0.0033 $ & $ 0.4919 \pm 0.0028 $ \\
1.0000 & $ 0.3583 \pm 0.0030 $ & $ 0.4433 \pm 0.0035 $ & $ 0.4979 \pm 0.0068 $ \\
\end{tabular}
\end{ruledtabular}
\end{table*}

\begin{table*}
\caption{ Peak amplitude of the 22 mode measured from the strain, news, and
$\Psi_4$.  Values and standard errors calculated in the same way as the peak frequency.
}
\label{tab:0ampvals}
\begin{ruledtabular}
\begin{tabular}{cccc}
$q$    & $r/m H_{22}^{\mathrm{peak}}$ & $r N_{22}^{\mathrm{peak}}$ & $rm \Psi_{4,22}^{\mathrm{peak}}$ \\
\hline
0.0100 & $ 0.0140 \pm 0.0000 $ & $ 0.0043 \pm 0.0000 $ & $ 0.0014 \pm 0.0000 $ \\
0.0667 & $ 0.0848 \pm 0.0003 $ & $ 0.0269 \pm 0.0001 $ & $ 0.0096 \pm 0.0001 $ \\
0.1000 & $ 0.1204 \pm 0.0004 $ & $ 0.0391 \pm 0.0002 $ & $ 0.0145 \pm 0.0001 $ \\
0.1667 & $ 0.1816 \pm 0.0009 $ & $ 0.0632 \pm 0.0014 $ & $ 0.0247 \pm 0.0007 $ \\
0.2000 & $ 0.2072 \pm 0.0009 $ & $ 0.0724 \pm 0.0003 $ & $ 0.0283 \pm 0.0002 $ \\
0.2500 & $ 0.2407 \pm 0.0010 $ & $ 0.0858 \pm 0.0003 $ & $ 0.0349 \pm 0.0001 $ \\
0.3333 & $ 0.2857 \pm 0.0010 $ & $ 0.1061 \pm 0.0003 $ & $ 0.0448 \pm 0.0001 $ \\
0.4000 & $ 0.3138 \pm 0.0010 $ & $ 0.1185 \pm 0.0004 $ & $ 0.0513 \pm 0.0003 $ \\
0.5000 & $ 0.3451 \pm 0.0009 $ & $ 0.1341 \pm 0.0003 $ & $ 0.0593 \pm 0.0002 $ \\
0.6000 & $ 0.3662 \pm 0.0010 $ & $ 0.1448 \pm 0.0004 $ & $ 0.0655 \pm 0.0002 $ \\
0.6667 & $ 0.3751 \pm 0.0012 $ & $ 0.1489 \pm 0.0006 $ & $ 0.0691 \pm 0.0009 $ \\
0.7500 & $ 0.3837 \pm 0.0011 $ & $ 0.1544 \pm 0.0004 $ & $ 0.0715 \pm 0.0004 $ \\
0.8500 & $ 0.3911 \pm 0.0011 $ & $ 0.1576 \pm 0.0004 $ & $ 0.0728 \pm 0.0002 $ \\
1.0000 & $ 0.3953 \pm 0.0018 $ & $ 0.1597 \pm 0.0007 $ & $ 0.0743 \pm 0.0004 $ \\
\end{tabular}
\end{ruledtabular}
\end{table*}

\subsection{Recoil velocities of non-spinning binaries}\label{sec:rvs0}

Consistent with our notation in Ref.~\cite{Healy:2014yta}, we expand the
non-spinning recoil as
\begin{equation}\label{eq:vm}
v_m=\eta^2 \delta m\left(A+B\,\delta m^2+C\,\delta{m}^4\right).
\end{equation}
where $\delta{m}=(m_1-m_2)/m$ and $m=(m_1+m_2)$ and $4\eta=1-\delta{m}^2$.

\begin{figure}
\includegraphics[angle=270,width=\columnwidth]{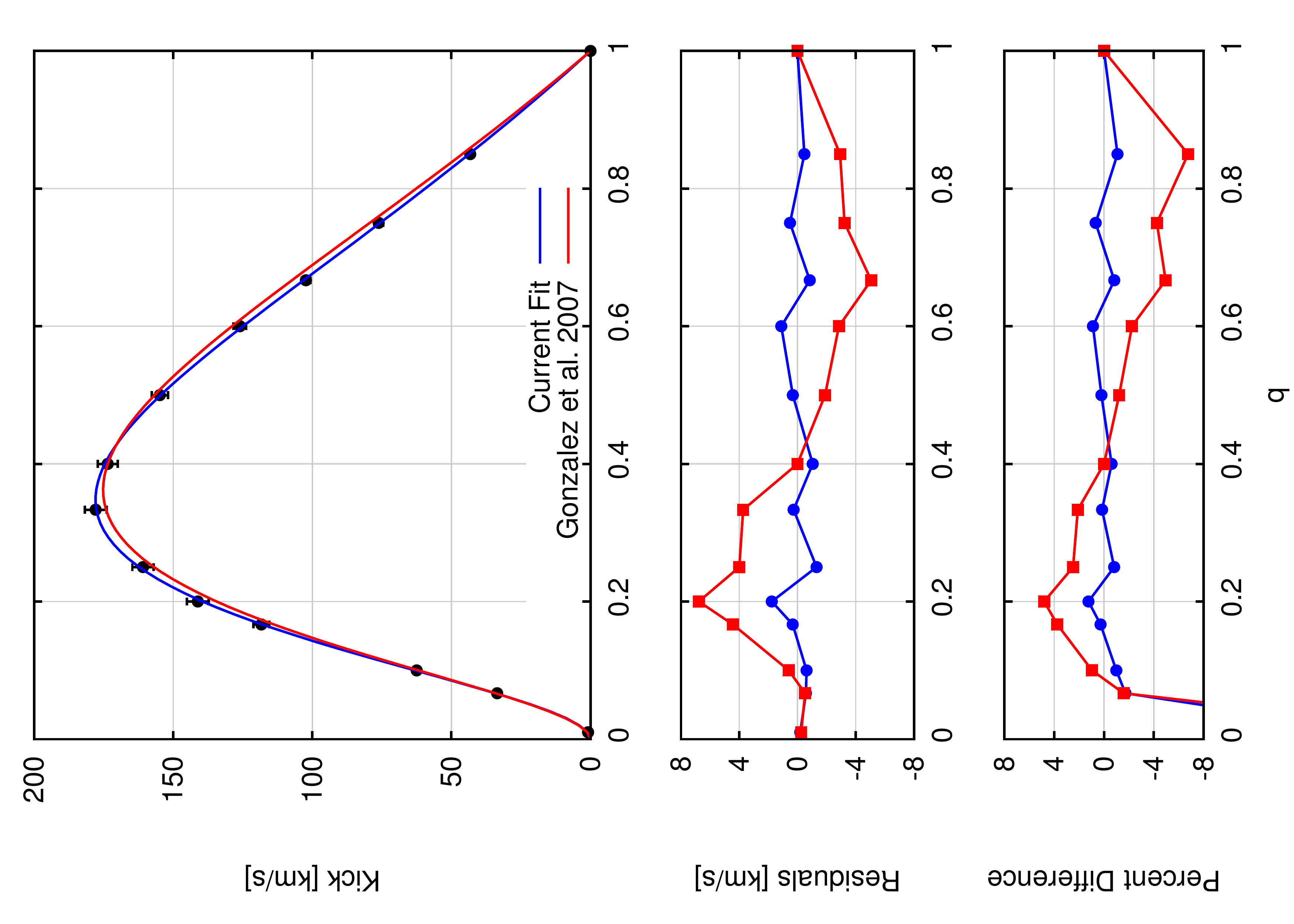}
\caption{Our current fit and the original Gonz\'alez {\it et al.}~\cite{Gonzalez:2007hi} fit to the recoils from nonspinning BHBs.
The panel below gives the residual and percent difference of both fits.
\label{fig:S0kicks}}
\end{figure}

The results of our runs
allow us to produce a new independent fit to the
non-spinning-black-hole-binary recoil. The new result and comparison
with the original fit of Gonz\'alez {\it et al.}~\cite{Gonzalez:2007hi}
(to independent data)
is displayed in Fig.~\ref{fig:S0kicks}.
We also display the residuals (of the order of $1$km/s) for the new
fit and compared to the corresponding (typically of several
km/s) deviations from the old fit.

Table \ref{tab:abc} gives the results of fitting the coefficients
$A$, $B$, and $C$ in
Eq.~(\ref{eq:vm}) to the 14 runs available here. We
find that the value of the additional parameter $C$ is
statistically significant and its inclusion improves the overall fit.
In addition we compare the old~\cite{Gonzalez:2007hi} $A,B$ parameters
with our fit to just these two parameters, i.e. setting $C=0$ and 
find that they are close but the differences are statistically significant.

\begin{table}
\caption{
Fitting to the recoil velocity of the remnant of nonspinning
black hole binaries by Eq.~(\ref{eq:vm}). Fit 2 only uses $A,B$ while
Fit 3 also fits $C$. Standard error for each fit is also given.}
\label{tab:abc}
\begin{ruledtabular}
\begin{tabular}{cccc}
Parameter & Fit~\cite{Gonzalez:2007hi} & Fit 2 & Fit 3 \\
\hline
$A$ & $-9210$ & $-8917\pm 73$ & $-8695\pm 53$ \\
$B$ & $-2790$ & $-4285\pm 261$ & $-6683\pm 424$ \\
$C$ & $0.0$   & $0.0$              & $4179\pm  702$\\
\end{tabular}
\end{ruledtabular}
\end{table}

We find that the maximum of the new fitting function lies at
$q=0.348$ with a recoil velocity of $178$km/s which shifts the maximum
to slightly lower mass ratio and slightly higher recoil velocity.  The 
Gonzalez~{\it et al.} fit finds a maximum recoil velocity of $175$km/s for
$q=0.362$.


\subsection{Peak luminosity of non-Spinning Binaries}\label{sec:pls0}

The formula to model the peak luminosity introduced in \cite{Healy:2016lce}
takes the following simple form for nonspinning binaries
\begin{equation}\label{eq:4plum}
L_{\rm peak} = (4\eta)^2\,\Big\{N_0 +
                     N_{2d}\,\dmt^2 +
                     N_{4f}\,\dmt^4\Big\}.
\end{equation}
Note that the radiated power in the particle limit scales as $\eta^2$
[see Ref.~\cite{Fujita:2014eta}, Eq. (16) and (20); evaluated at the ISCO
for its peak value].

\begin{figure}
\includegraphics[angle=270,width=\columnwidth]{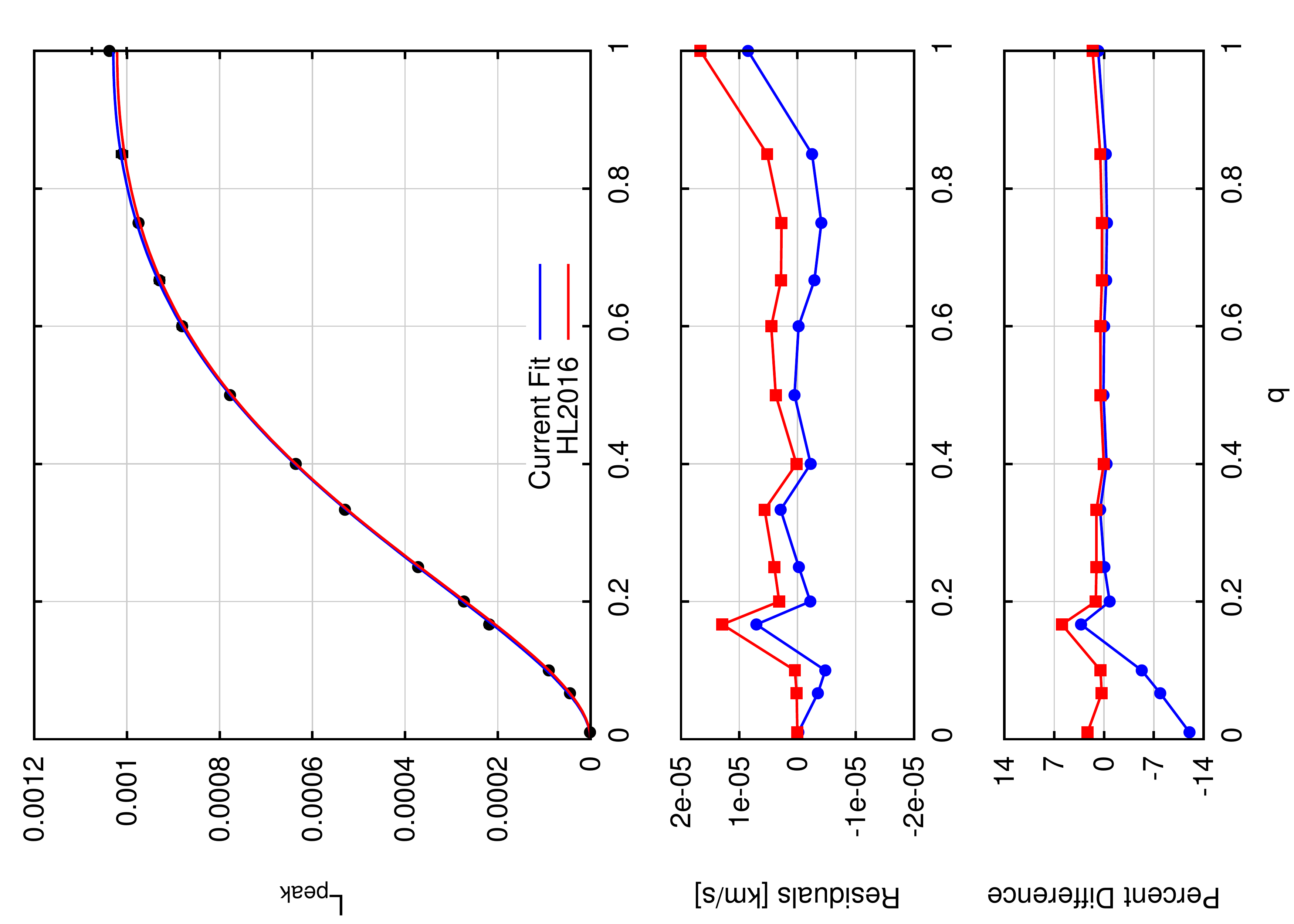}
\caption{Our current fit and previous fit \cite{Healy:2016lce} to the peak luminosity
from nonspinning BHBs.
\label{fig:S0pLum}}
\end{figure}

The results of fitting the parameters $N_0$, $N_{2d}$, and $N_{4f}$ to the
peak luminosity of our 14 simulations is displayed in Fig.~\ref{fig:S0pLum}
and compared to the previous fit in Ref.~\cite{Healy:2016lce} (note
that \cite{Healy:2016lce}  included
spinning and nonspinning simulations to determine the fitting
parameters). We summarize the results in 
Table~\ref{tab:plum}.

The results of this comparison is again a reduction of the residuals
over the mass ratio range studied here and provides new values to the
fitting parameters $N_0,N_{2d},N_{4f}$ to be used in future hierarchical
approaches to formulate the modeling of the more general case of spinning
precessing black hole binaries. Note that increasing the resolution
leads to an increase in the peak luminosity (likely due to the
decreased effects of artificial dissipation at high resolution)
 This is reflected in 
the residuals over the whole range of the mass ratios,
$q=1$ to $q=1/100$. 
The peak luminosity values reach a maximum for equal mass binaries,
producing a peak just above $10^{-3}$ in dimensionless units and
vanishing in the particle limit as $\eta^2$.

\begin{table}
\caption{
The fitting coefficients for the peak luminosity Eq.~(\ref{eq:4plum}).
}
\label{tab:plum}
\begin{ruledtabular}
\begin{tabular}{cr}
Parameter & $L_{peak}$\\
\hline
$N_0$    & $\ 1.029\times10^{-3}\pm2.454\times10^{-6}$\\
$N_{2d}$ & $-4.474\times10^{-4}\pm4.045\times10^{-5}$\\
$N_{4f}$ & $\ 3.086\times10^{-4}\pm9.310\times10^{-5}$\\
\hline
\end{tabular}
\end{ruledtabular}
\end{table}

\subsection{Peak frequency and amplitude of non-Spinning Binaries}\label{sec:oms0}

Analogously to the previous formula to model the peak luminosity,
we introduce the following fitting formula for the peak frequency
of the $(2,2)$ mode of the gravitational wave strain for 
nonspinning binaries
\begin{equation}\label{eq:4om}
m\omega_{22}^{\mathrm{peak}} = \Big\{W_0 +
                     W_{2}\,\dmt^2 +
                     W_{4}\,\dmt^4\Big\},
\end{equation}

\begin{figure}
\includegraphics[angle=270,width=\columnwidth]{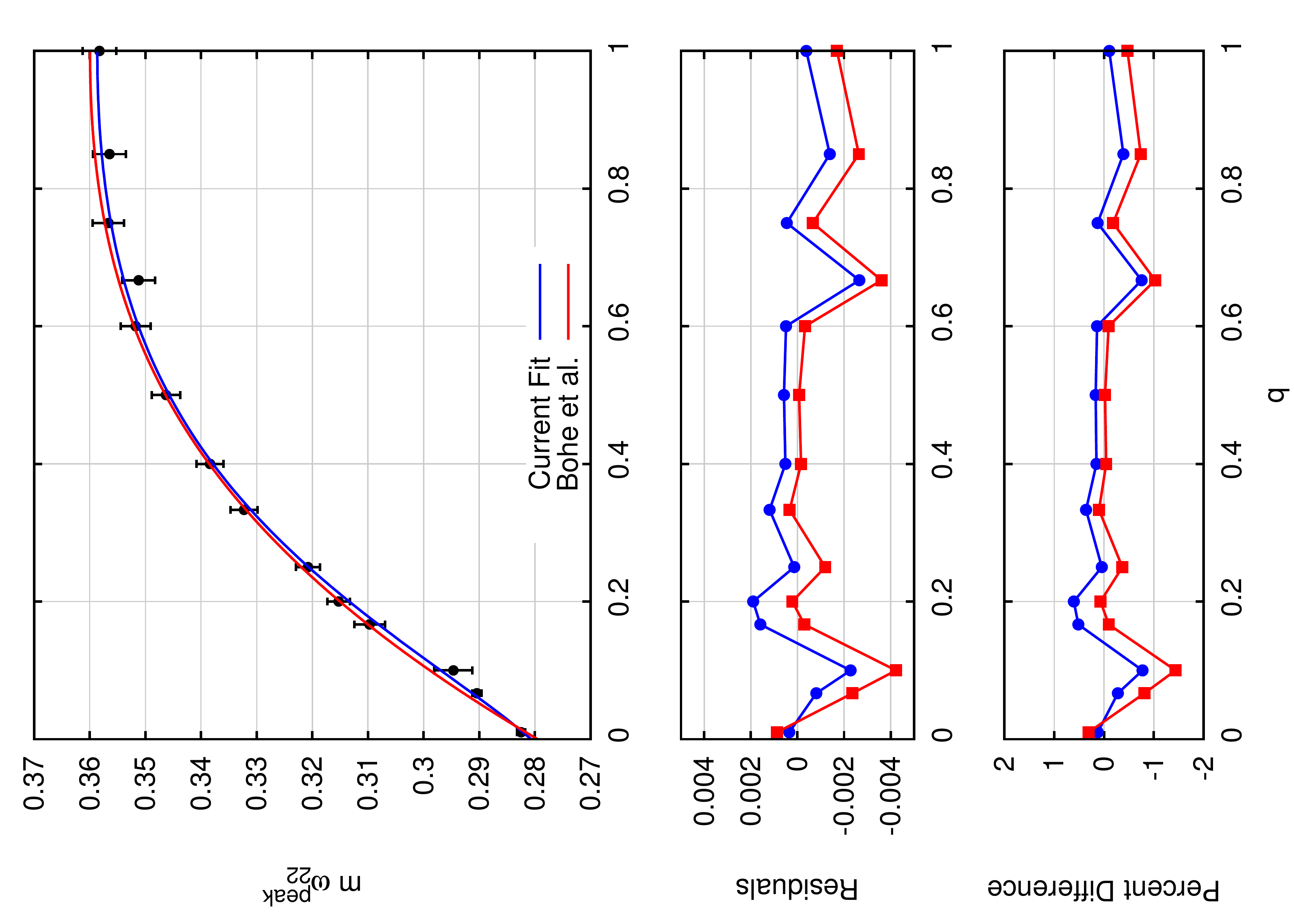}
\caption{Current fit and the Boh\'e {\it et al.}~\cite{Bohe:2016gbl}
fit to the peak waveform frequency from nonspinning BHBs.
\label{fig:S0om}}
\end{figure}

The results of fitting the parameters $W_0$, $W_{2}$, and $W_{4}$ to the
peak frequency of our 14 simulations are given in Table \ref{tab:omfit}
and are displayed in Fig.~\ref{fig:S0om}. 
We note here that in the $q\to0$ limit, the frequency
approaches a value of $\approx0.2807$, which is close to 
the particle limit $0.2795$ reported in the \cite{Bohe:2016gbl},
Eq. (A6) [and to (twice) the frequency of the ``ibco'', $0.25$
that innermost bounded
circular orbit for nonspinning black holes \cite{Bardeen:1972fi}].
While towards the equal-mass limit the frequency increases to
$W_0\sim0.358$. Note that~\cite{Bohe:2016gbl} [Eq. (A7)] finds
a peak frequency of $0.36$ in the equal-mass limit.

\begin{table}
\caption{
Fitting to the peak frequency of the 22 mode of the strain produced by
black hole binaries by Eq.~(\ref{eq:4om}) and Eq.~(\ref{eq:4omp}).
 Standard error for each fit is also given.}
\label{tab:omfit}
\begin{ruledtabular}
\begin{tabular}{crcr}
Parameter & Fit 1& Parameter &Fit 2\\
\hline
$W_0$ & $ 0.3586\pm0.0008$ & $W'_0$ & $ 0.3579\pm0.0011$ \\
$W_2$ & $-0.1210\pm0.0037$ & $W'_2$ & $ 0.2471\pm0.0094$ \\
$W_4$ & $ 0.0431\pm0.0034$ & $W'_4$ & $ 0.2713\pm0.0129$ \\
\end{tabular}
\end{ruledtabular}
\end{table}

If we impose the particle limit peak frequency, $m_f\,\Omega_p=0.2795$ into our formula,
we have the alternative Fit 2:

\begin{equation}\label{eq:4omp}
m\omega_{22}^{\mathrm{peak}} = (4\eta)\, \Big\{W'_0 +
                     W'_{2}\,\dmt^2 +
                     W'_{4}\,\dmt^4\Big\} +
                     m_f\,\Omega_p\,\dmt^6,
\end{equation}
where $\eta=(1-\delta{m}^2)/4$.

Note also that the peak frequency for the Weyl scalar $\psi_4$
(instead of the strain $h$ studied here), was studied in
\cite{Healy:2014eua}
in connection with the quasinormal modes of the final remnant
and a fitting to the peak frequency produced by numerical simulations
was used to calibrate EOB models in \cite{Taracchini:2012ig}.

\begin{table}
\caption{
Fitting to the peak amplitude of the 22 mode of the strain produced by
black hole binaries by Eq.~(\ref{eq:4ph}).
 Standard error for each fit is also given.}\label{tab:fit_amp}
\label{tab:ampfit}
\begin{ruledtabular}
\begin{tabular}{cr}
Parameter & $r/m H_{22}^{peak}$ \\
\hline
$H_0$ & $ 0.3980\pm0.0003$ \\
$H_2$ & $-0.0558\pm0.0019$ \\
$H_4$ & $ 0.0183\pm0.0019$ \\
\end{tabular}
\end{ruledtabular}
\end{table}

\begin{figure}
\includegraphics[angle=270,width=\columnwidth]{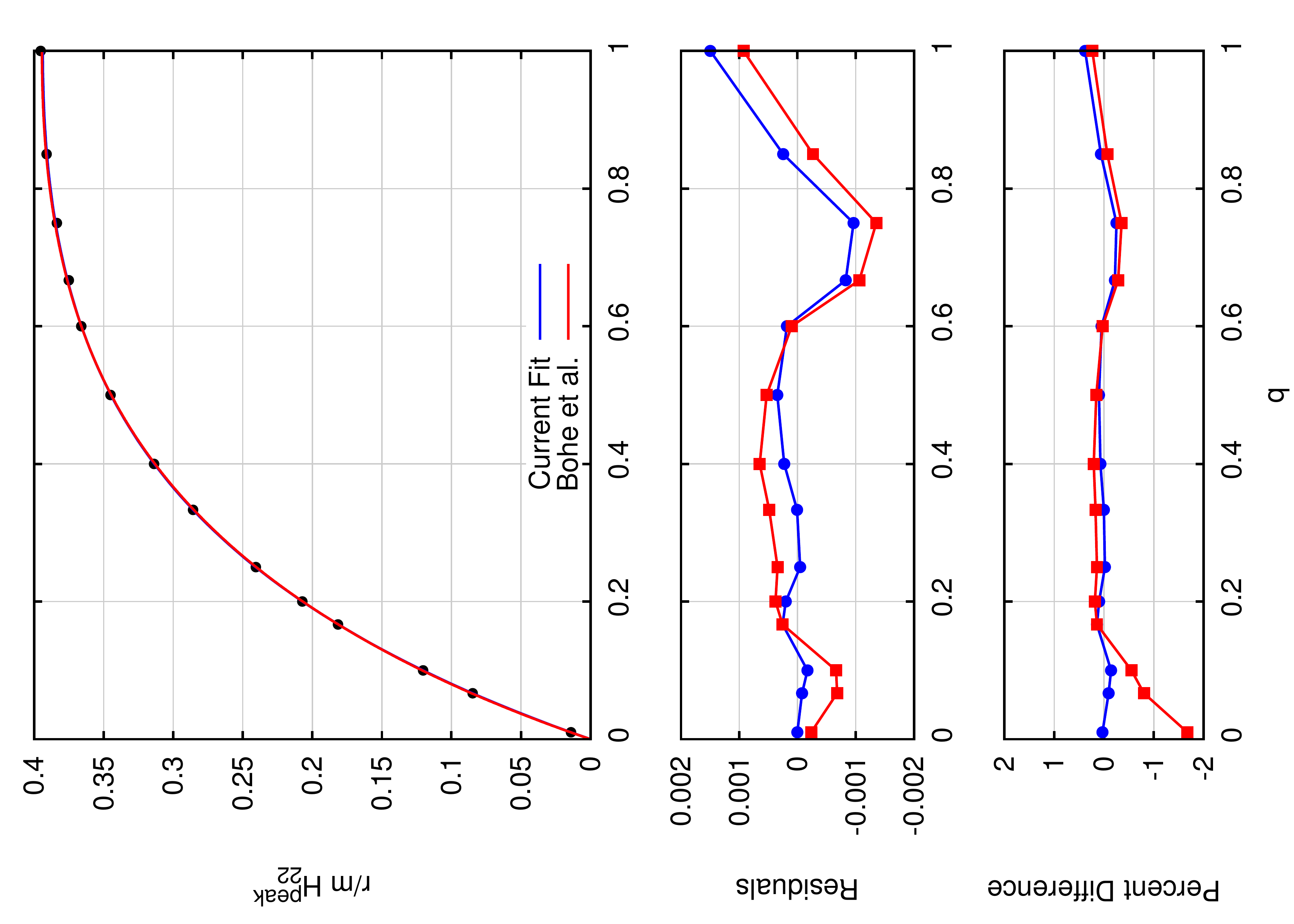}
\caption{Current fit and the Boh\'e {\it et al.}~\cite{Bohe:2016gbl}
fit to the peak strain amplitude from nonspinning
  BHBs. 
\label{fig:S0amp}}
\end{figure}

In addition to modeling the peak frequency, We also model the peak
amplitude (of the strain $h$) from the merger of nonspinning
 binaries using the expansion
\begin{equation}\label{eq:4ph}
h_{\rm peak} = (4\eta)\,\Big\{H_0 +
                     H_{2}\,\dmt^2 +
                     H_{4}\,\dmt^4\Big\}.
\end{equation}
The results from this fit are summarized in Table~\ref{tab:fit_amp} and
Fig.~\ref{fig:S0amp}.

\section{Conclusions and Discussion}\label{sec:Discussion}

The study of remnant formulas has been of interest since the pioneering
work using the Lazarus approach \cite{Baker:2003ds, Campanelli:2004zw} 
over a decade ago.
The breakthroughs in numerical relativity allowed for a more complete
study and a number of increasingly general and accurate phenomenological
formulas have been put forward over the years (See for instance
\cite{Lousto:2009mf,Lousto:2013wta,Hofmann:2016yih,Jimenez-Forteza:2016oae}
and references therein).
The first detection of gravitational waves from the merger of two
black holes \cite{Abbott:2016blz} produced a renewed interest in the remnant formulas 
\cite{TheLIGOScientific:2016wfe,TheLIGOScientific:2016src,Abbott:2016apu,Abbott:2016nmj,TheLIGOScientific:2016pea}. 

The remnant formulas for the final mass and spin of the product of
two merged black holes can be made very accurately since we can compute
the final masses and spins (magnitudes) from the
isolated horizon formulas \cite{Dreyer02a}.
Alternatively, one can compute
those quantities from the energy and angular momentum carried out to infinity
by the waveforms and subtract those values from the initial total mass
and angular momentum of the system. This method, provides a consistency
check to the isolated horizon computation, but requires higher resolutions
to achieve comparable accuracy (See appendices in Refs. 
\cite{Healy:2014yta,Healy:2016lce}).
A third method can be also used by measuring directly the quasinormal
modes in the late ringdown phase of the waveform and relate them
to the mass and spin of a perturbed Kerr black hole 
(See for instance Table III in \cite{Dain:2008ck} and references therein).

The recoil velocity of the remnant and the peak luminosity of merging
binary black holes is also of renewed interest
\cite{Keitel:2016krm,Healy:2016lce}
but those quantities (as well as the peak frequency and amplitude) 
are computed from the waveforms 
(but see~\cite{Krishnan:2007pu}) 
and hence are computed with less accuracy in the
routine simulations that do not reach ultra-high resolutions.
In this paper, we have revisited the study of nonspinning binaries
with a set of three resolutions (low, medium, high) that allows us to 
confirm that we are in the convergence regime and that we are able to extrapolate
to infinite resolution to obtain a more accurate recoil and peak
luminosity than by the standard runs \cite{Healy:2016lce}.
This serves to establish a new set of fitting coefficients that, in a 
hierarchical approach, will serve as fixed constants in 
the new fittings (or refitting) of the more general remnant formulas
(for the spinning \cite{Healy:2016lce} and precessing \cite{Zlochower:2015wga}
 binaries).
We have also introduced formulas for the gravitational wave frequency and
amplitude at the peak of the strain. This provides further information about
the full numerical simulations that can be used  \cite{Bohe:2016gbl}
to improve the approximate modeling of gravitational waveforms
used for data analysis of gravitational wave signals measured by
laser interferometric detectors.
The fittings (\ref{eq:4plum})-(\ref{eq:4ph}) can also be used
for a consistency test of general relativity
\footnote{We thank M.Campanelli for making this point.}
by comparing the prediction
of the peak luminosity/amplitude and the frequency of this peak from the above
formulas (and its generalization to spinning black holes)
with an actual measurement from a gravitational wave signal
\cite{Cornish:2014kda,Klimenko:2015ypf,Lynch:2015yin,Becsy:2016ofp,TheLIGOScientific:2016uux}.

\acknowledgments 
The authors thank M. Campanelli, D.Keitel, N.K.J-McDaniel, H. Nakano, and R. O'Shaughnessy for discussions on this work.
The authors gratefully acknowledge the NSF for financial support from Grants
No.  PHY-1607520, No. ACI-1550436, No. AST-1516150, and No. ACI-1516125.
Computational resources were provided by XSEDE allocation
TG-PHY060027N, and by NewHorizons and BlueSky Clusters 
at Rochester Institute of Technology, which were supported
by NSF grant No. PHY-0722703, DMS-0820923, AST-1028087, and PHY-1229173.
This research was also part of the Blue Waters sustained-petascale computing
NSF projects ACI-0832606, ACI-1238993, and OCI-1515969, OCI-0725070. 


\bibliographystyle{apsrev4-1}
\bibliography{../../Bibtex/references}

\end{document}